\documentstyle[preprint,aps]{revtex}
\begin{document}
\draft
\def \beq{\begin{equation}}
\def \eeq{\end{equation}}
\def \beqarr{\begin{eqnarray}}
\def \eeqarr{\end{eqnarray}}

\title{Coulomb and Hard Core Skyrmion Tails}

\author{E. H. Rezayi$^1$ and S. L. Sondhi$^2$.}

\address{
$^1$Department of Physics,
California State University
Los Angeles, CA 90032, USA
}

\address{
$^2$Department of Physics,
Princeton University,
Princeton, NJ 08544, USA
}

\date{\today}
\maketitle
\begin{abstract}
Quantum Hall skyrmions are quantized solitons of
a ferromagnetic $O(3)$ $\sigma-$model.
The reference, classical, solutions depend upon
the interaction between the electrons and exhibit
completely different asymptotic profiles for the
physical Coulomb interaction than for the hard core 
interaction frequently used to generate variational 
wavefunctions. In this note we show, by means of
numerical calculations on (large) finite size systems
at $\nu=1$, that this physically important difference,
crucial for a sharp definition of their statistics, 
persists for the quantized skyrmions at $\nu=1$.
\end{abstract}

Theoretical and experimental work over the last few years has adduced
strong evidence that the physics of the quantum Hall (QH) system in the
vicinity of (filling factor) $\nu=1$ is best understood in terms
of a ferromagnetic ground state right at commensuration which supports
charged excitations with non-trivial spin topology christened 
``skyrmions'' \cite{sondhi1,hf,expts}. The skyrmions arise as quantized
solitons in a two dimensional $O(3)$ $\sigma-$model which describes the 
low energy physics of the $\nu=1$ state, and {\it mutatis mutandis} of 
other ferromagnetic QH states. Very general considerations of homotopy
suggest the possibility of topologically stable and non-trivial classical
solitons in any two dimensional $O(3)$ $\sigma-$model \cite{rajaraman}
but the issue of their dynamical stability is more complicated. As
was first noted by Belavin and Polyakov \cite{bp}, the leading (gradient 
squared) term in their energy functional is scale invariant in two dimensions;
hence the explicit solutions obtained by them exhibit an arbitrary scale 
parameter and algebraic asymptotic profiles. Consequently, dynamical
stability, i.e. the existence of a well defined size for the skyrmions,
depends crucially on the sub-leading terms which are system specific. 

In the theory of the $\nu=1$ QH this issue can be resolved in two very
different ways. For the physical, Coulomb, interaction, the effective
$\sigma-$model written in terms of the order parameter field ${\bf n}({\bf r})$
is modified to \cite{sondhi1}
\begin{eqnarray}
\label{effective}
{\cal L}_{eff} = \frac 1 2 {\bar \rho} {\bf {\cal A}}({\bf n}) \cdot 
\partial_t {\bf n}
- \frac{1}{2}\rho^s (\nabla {\bf n})^2
+ \frac 1 2 g \overline{\rho} \mu_B {\bf n} \cdot {\bf B}
-\frac{e^2}{2 \epsilon} \int d^2r' \frac{q({\bf r}) 
q({\bf r}')} {|{\bf r}-{\bf r}'|} \ \ \ ,
\end{eqnarray}
neglecting a Hopf term that does not affect the structure of the skyrmions.
Here ${\bf {\cal A}}$ is the vector potential of a unit magnetic monopole, 
$\rho^s$ the spin stiffness ($\rho^s = e^2/(16\sqrt{2 \pi}
\epsilon \ell)$ for $\nu=1$), $\epsilon$ the semiconductor dielectric 
constant, $\ell$ the magnetic length and $q({\bf r})= {\bf n} \cdot 
(\partial_x {\bf n} \times \partial_y {\bf n})/4 \pi$ is the topological 
density of the spin field.
The Zeeman (third) and Coulomb (fourth) terms in  ${\cal L}_{eff}$ break the 
spatial scale invariance of the gradient (second) term and their competition 
then sets the  size and energy of the skyrmions which now depend on the 
dimensionless ratio $\tilde g = (g\mu_B B)/(e^2/ \epsilon \ell ^2)$ of the 
Zeeman energy to the Coulomb energy.  
On the other hand, one can study a ``hard core'' interaction, consisting of 
a delta function projected to the lowest Landau level, which is a standard
device in QH theory to generate variational wavefunctions \cite{haldane}. 
Remarkably, the effective $\sigma-$model for this problem contains {\it
exactly} the first three terms, i.e. it supports skyrmions of 
arbitrary size $\lambda$ at $g \mu_B B =0$ given by the Belavin/Polyakov 
solutions:
\begin{equation}
  n_x =  \sqrt{1 - f^2(r)} \, \cos(\theta) \ , \ \ \
  n_y =  \sqrt{1 - f^2(r)} \, \sin(\theta) \ , \ \ \
  n_z = f(r)  \ \ \ , 
\label{ansatz}
\end{equation}
where $f(r)=[(r/\lambda)^{2}-4]/[(r/\lambda)^{2}+4]$ \cite{fn1}. For 
$g \mu_B B \ne 0$, the stable solution has vanishing size. 
Despite this last pathology of the hard core problem, it would appear 
that its solutions could again be used for a variational solution of 
the Coulomb problem, i.e. by optimizing $\lambda$.

It is a fact of some consequence, that this cannot be done. It turns out
\cite{sondhi1,lejnell} that the skyrmion profile in Eq (\ref{ansatz}) decays
too slowly ($n^z -1 \sim (\lambda/r)^2$) to yield a finite answer for the 
Zeeman energy, necessitating
a calculation of its true asymptotic form in the Coulomb problem which
can be shown to go as $n^z -1 \sim e^{-r/\sqrt{g}}$, i.e. the Coulomb
skyrmion has a much shorter tail than the hard core variant. Clearly, if this
was {\it not} taken into account one would falsely conclude that Coulomb
skyrmions are always microscopic. Further, it was shown by Yang and Sondhi
\cite{ky-sls} that the statistics of the skyrmions is ill-defined in hard core
multi-skyrmion states precisely on account of the long tails of the skyrmions
but well defined for the shorter tailed Coulomb versions. Finally, the 
difference in asymptotic behavior was shown recently to lead to binding between
singly charged skyrmions at small Zeeman energies, albeit at unphysically small
values \cite{daniel}.

The above considerations were all in the framework of classical solutions
of the $\sigma-$model and the closely related Hartree-Fock 
approximation \cite{hf}. 
The corresponding quantum states require inclusion of fluctuations that 
quantize spin and angular momentum \cite{nw,mfb} and restore the
symmetries broken in the classical solutions, and one may wonder if the 
distinction between the tails of hard core and Coulomb skyrmions survives 
when this is done. In this note we report numerical calculations on the 
smallest skyrmions, i.e. those with one extra reversed spin for which the 
fluctuations are the most severe, and show that this important distinction 
is indeed preserved. 

\noindent
{\bf Results:} We study a QH system on a sphere \cite{haldane} at one flux 
quantum
less than the finite size version of $\nu=1$, i.e. with $N_\phi$ flux quanta
passing through the surface of the sphere and $N=N_\phi \equiv 2K$ particles. 
This is the one quasielectron sector and its states are related to those in 
the one quasihole sector ($N=N_\phi +2$) by a particle-hole transformation.
The spectrum in this sector has a set of states with equal values of the
angular momentum $L$ and spin $S$. The states with $L=S=N/2-1$ are the
multiplet of the fully polarized quasielectron (it has just one reversed
spin) and those with lower $L,S$ are skyrmions of various sizes culminating
in the infinite skyrmion (it has the size of the system) with $L=S=0$. We have
studied the smallest of the skyrmion states, i.e. with $L=S=N/2-2$ and maximum
$L_z$ for simplicity, for system sizes up to 300 electrons  for both hard core
and Coulomb interactions.

For the hard core interaction, an exact wavefunction can be written down. In
second quantized notation this takes the form,
\begin{equation}
|\Psi\rangle = \frac{1}{\sqrt{A}} \sum_m  \{ \frac{1}{\sqrt{2K}} 
c_{m\uparrow} c^\dagger_{m \downarrow} c^\dagger_{K-1,\downarrow} +
\sqrt{\frac{K+m}{K-m}} c_{m\uparrow} c^\dagger_{m-1 \downarrow} 
c^\dagger_{K,\downarrow} \} |\nu_{\uparrow}=1\rangle \ \ \ ,
\end{equation}
where $A=(2K+1) \sum_{n=2}^{2K} 1/n\approx(2K+1)(\ln(2K)+\gamma-1)$, 
and $\gamma$ is the Euler's constant. From this we can easily compute
the occupation of various orbitals by the down spins,
\begin{eqnarray}
n(K) &=& 1 -(1-1/N_\phi)/A  \nonumber \\
n(K-1) &=& N_\phi(1-1/N_\phi)/A \nonumber \\
\vdots \\
n(K-m) &=& (N_\phi+1)/(mA) -(1 -1/N_\phi)/A \ \ \ m\ge 2 .
\end{eqnarray}
It is clear from these formulas that one of the down spins is localized at the
north pole (top orbital).  For N=300, in Fig~1 we plot the spatial density 
for the 
other down spin
$ \sigma_{\downarrow}(r)={1\over R^2 }\sum_{m=-K}^{K-1}|\psi_m(\hat\Omega)|^2 
n_{\downarrow}(m)$ by 
excluding the $m=K$ orbital.  Here $r=2R\sin{\theta/2}$ is the chord distance,
$R=\sqrt{K}\ell$ is the radius of the sphere.
Evidently, $\sigma_{\downarrow}(r)$ is peaked near the nominal
location of the quasielectron. In the same figure we plot the
result for the exact wavefunction for Coulomb interactions which
was obtained by numerical diagonalization.  One can
see a considerable difference in the degree of localization of the second
reversed spin between these two potentials.  The difference in the tails,
however, may not be
immediately
apparent from this plot. It becomes apparent when we plot
the data on a double-logarithmic scale in Fig~2, where it is 
clearly seen that the down spins are strongly localized in the
Coulomb case when compared with the hard core wavefunction; at
the opposite pole there are seven orders of magnitude separating
the two results($3.1\times10^{-14}$, vs. $8.5\times10^{-6}$)! 
Moreover, the hard core distribution follows
the $1/r^2$ decay expected from the sigma model analysis while
the Coulomb result decays faster than any power consistent
with the predicted exponential behavior.  The asymptotic behavior on
the sphere for the hard core potential to leading order in system size
can easily shown to be ${N_\phi+1\over 4\pi\ell^2A}
({4\over (r/\ell)^2}-{1\over K})$, see the dashed line in Fig~1.
Yet another probe
of this difference is the large system behavior of the down
spin occupation of a {\em given} orbital near the location
of the quasielectron. The hard core wavefunction is easily
seen to lead to a logarithmic {\em vanishing} of this quantity 
in the large system limit which implies that the quasielectron
is not a well defined object at infinite system size: one of
its reversed spins is spread over the entire system. By contrast
the Coulomb quasielectron is well defined, examination of the
variation of the occupation with system size displays clear
evidence of finiteness in the large system limit (Fig~3). For comparison,
we have produced a similar plot for the hard-core potential in Fig~4.  Only
the occupation of the top orbital shows scaling with system size consistent
with its being localized.  Others show noticeable deviations as expected for
delocalization of the second reversed spin.

In summary, there is a crucial difference in the asymptotic behavior
of hard core (scale invariant $\sigma-$model) skyrmions and
Coulombic skyrmions which is present also in the fully quantized
and symmetric many-body states of the lowest Landau level problem.
Hard core skyrmions have algebraic profiles and are ill-defined 
in the large system limit while Coulomb skyrmions have exponential
profiles and have a definite size. This difference is crucial to
the energetic stability of Coulomb skyrmions, their possessing
a well defined statistics and to their binding at extremely small 
Zeeman energies.  Two remarks are in order. First, the distinction we have 
drawn here between the skyrmions of the hard-core and the Coulomb potentials
trivially vanishes for infinite skyrmions as they have no tails.
Second, {\em any} generic perturbation of the hard core interaction (one
which produces a direct interaction in addition to the exchange) will
break the scale invariance and give a well defined size to the skyrmions;
the hard core limit, though important in that it enables an exact
analysis, is non-generic in this regard.  

\acknowledgements
We are grateful to A. Karlhede, S. A. Kivelson and K. Yang for useful
discussions. This work was supported in part by NSF grant No. DMR-9632690
and the A. P. Sloan Foundation (SLS) and NSF grant No. DMR-9420560 (EHR).

\begin{figure}
\caption{The spatial distribution of the second reversed spin for the Coulomb 
and the hard core potential for a 300-electron size systems.  The dashed line 
is the asymptotic behavior to leading order in system size given in text.
The inset shows the tails of the distribution.} 
\bigskip

\caption{The log-log plot of Fig~1.  The differences in the tail can be
clearly seen.  The density for the Coulomb tail is curved indicating it
vanishes faster than any power. The straight line segment for hard core gives
the $1/r^2$ dependence.}
\bigskip

\caption{The plot of occupation amplitudes for the top 4 orbitals for the 
Coulomb potential for up to N=300 plotted versus 1/N.  All are essentially
independent of system size and indicate localization of the second 
reversed spin.}
\bigskip

\caption{Same as Fig~3 except for hard core potential plotted for comparison.
Only the first reversed spin at the top orbital can be seen to be localized.
The others show a noticeable decrease  and ultimately
vanish logarithmically with system size.}

\label{profiles}
\end{figure}


\begin{references}
\bibitem{fn1} This conclusion followed from the work of one of us (EHR)
in \cite{sondhi1} that showed that the excitations of the hard
core model were degenerate in energy at different sizes; we do not know
of a constructive proof of this at present. This connection was noted
in the context of the Hartree-Fock approximation in \cite{hf}.
\end{references}
\end{document}